\documentclass[prl,twocolumn,showpacs,letterpaper,showpacs,superscriptaddress]{revtex4}
\usepackage{graphicx,amsmath,amssymb,amsfonts,latexsym,color,dcolumn,bm,epsfig,subfigure}

\renewcommand{\imath}[0]{\mathrm{i}}

\newcommand{\mathbfh}[1]{\hat{\mathbf{#1}}}

\begin{document}

\title{Quantum friction and fluctuation theorems} 

\author{F. Intravaia}
\affiliation{Theoretical Division, MS B213, Los Alamos National Laboratory, Los Alamos, New Mexico 87545, USA}
\affiliation{Max-Born-Institut, 12489 Berlin, Germany}
\author{R. O. Behunin}
\affiliation{Theoretical Division, MS B213, Los Alamos National Laboratory, Los Alamos, New Mexico 87545, USA}
\affiliation{Center for Nonlinear Studies, Los Alamos National Laboratory, Los Alamos, New Mexico 87545, USA}
\affiliation{Department of Applied Physics, Yale University, New Haven, Connecticut 06511, USA}
\author{D. A. R. Dalvit}
\affiliation{Theoretical Division, MS B213, Los Alamos National Laboratory, Los Alamos, New Mexico 87545, USA}

\date{\today}

\begin{abstract} 
We use general concepts of statistical mechanics to compute the quantum frictional force on an atom moving at constant velocity above a planar surface. We derive the zero-temperature frictional force using a non-equilibrium fluctuation-dissipation relation,
and show that in the large-time, steady-state regime quantum friction scales as the cubic power of the atom's velocity. We also discuss
how approaches based on Wigner-Weisskopf and quantum regression approximations fail to predict the correct steady-state zero temperature
frictional force, mainly due to the low frequency nature of quantum friction. 
\end{abstract}

\maketitle


A remarkable example of fluctuation-induced interactions is quantum friction, the drag force experienced between two bodies in relative motion in vacuum, associated with the energy and momentum transfer from one body to the other mediated by the quantum electromagnetic field.
Radiation mediated friction is deeply rooted in the foundations of quantum mechanics and it was already discussed by Einstein in his seminal 1917  paper on black body spectrum \cite{Einstein17}. Quantum friction has recently attracted attention in the context of macroscopic bodies and atoms in linear \cite{Volokitin07} or rotational \cite{Zhao12} motion above a surface, Coulomb drag in electron transport phenomena
\cite{Volokitin11}, and as the dissipative counterpart of the dynamical Casimir effect \cite{Dalvit00}. Several authors  \cite{Mahanty80,Schaich81,Tomassone97,Volokitin02,Dedkov02,Zurita-Sanchez04,Scheel09,Barton10b,Pieplow13} have obtained quite diverse results for the atom-surface drag at zero temperature, making different predictions as to its dependence on the velocity of the atom and the atom-surface separation. 
Here we revisit the problem of quantum friction using general concepts of
quantum statistical mechanics. We derive a quantum non-equilibrium fluctuation-dissipation theorem (FDT) for an atom in steady-state motion above a surface and compare its predictions with the quantum regression theorem (QRT).
  
We first consider the 
prototype problem of a static atom above a planar material surface at zero temperature. The atom is described by an electric dipole operator  
$\hat{\bf d}$ located at position ${\bf r}_a$. In a simple two-state system model (ground $|g\rangle$ and excited state $|e\rangle$) the atomic electric dipole operator is
given by $\hat{\bf d} = {\bf d} \hat{\sigma}_{1}$, where ${\bf d}$ is the (real) dipole vector and $\hat{\sigma}_{1}=|e\rangle\langle g|+|g\rangle\langle e|$ describes the internal degrees of freedom \cite{Allen75} (the generalization to multi-level atoms is straightforward
\cite{Buhmann04,Scheel09}). Alternatively,
in a model of the atom as a harmonic oscillator, $\hat{\bf d} = {\bf d} \hat{q}$, where $\hat{q}$ is a dimensionless position operator \cite{hamiltonian}. At any given time $t$, the force on the atom normal to the surface is given by $F_z(t) = \langle \hat{\bf d}(t) \cdot \partial_{z_a} \hat{\bf E}({\bf r}_a,t) \rangle$. From the Maxwell equations the electric field operator can be written as
$\hat{\bf E}({\bf r},t) =\hat{\bf E}_{0}^{(+)}({\bf r},t)+ (i/\pi) \int_{0}^{\infty} d\omega\, \int_{0}^{t}d\tau e^{-\imath \omega\tau}\underline{G}_{I}({\bf r},{\bf r}_a, \omega)\cdot \hat{\bf d}(t-\tau) + h.c.$,
where $\underline{G}$ is the electric Green tensor of the surface (the subscripts $R$ and $I$ will denote real and imaginary part), and $\hat{\bf E}_0^{(+)}$ denotes the positive-frequency solution for the electric field in the absence of the atom. We will assume 
that the initial atom + field/matter state is factorizable, $\hat{\rho}(0)=\hat{\rho}_{\rm a}(0)\bigotimes \hat{\rho}_{\rm fm}(0)$, with the joint field/matter subsystem in its vacuum state. Using normal ordering the force can be written as
\begin{eqnarray}
F_z(t) &=&
{\rm Re} \left\{ \frac{2 i}{\pi}\int_{0}^{\infty}d\omega\, \int_{0}^{t}d\tau e^{-i \omega \tau}  \right. 
\label{CPforce} \\
&& \times \left. {\rm Tr} \left[\langle \hat{\bf d}(t)  \hat{\bf d}(t-\tau)\rangle\cdot \partial_{z_a}\underline{G}_{I}(\mathbf{r}_{a},\mathbf{r},\omega)
|_{{\bf r}={\bf r}_a} \right] \right\} , \nonumber
\end{eqnarray}
where the trace is over the vector coordinates and  $\langle \ldots \rangle$ denotes expectation value over the initial state. 
Note that in this equation $\hat{\bf d}(t)$ represents the exact dynamics of the dipole operator, including back action from the field/matter.
The two-time correlation tensor $\underline{C}_{ij}(t,t-\tau) \equiv \langle \hat{\bf d}_i(t)  \hat{\bf d}_j(t-\tau)\rangle$ will be a key quantity in what follows.  For 
the equilibrium problem being considered, the stationary ($t \rightarrow \infty$) density matrix of the coupled atom-field-matter system has the Kubo-Martin-Schwinger (KMS) form, $\hat{\rho}(\infty) =\hat{\rho}_{\rm KMS} \propto e^{-\beta \hat{H}}$ ($\beta$ is the inverse temperature and $\hat{H}$ is the system's Hamiltonian); at zero temperature $\hat{\rho}(\infty)$ corresponds to the ground state of the whole system. Hence, in the stationary state the two-time correlation tensor tends to  $ \underline{C}_{ij}(\tau) \equiv  {\rm tr} \{ \hat{\bf d}_i(\tau) \hat{\bf d}_j(0) \hat{\rho}_{\rm KMS} \}$, and the zero-temperature FDT \cite{Callen51} relates the corresponding power spectrum  $\underline{S}(\omega) = (2 \pi)^{-1} \int_{-\infty}^{\infty} d\tau e^{i \omega \tau} \underline{C}(\tau)$ with the atom's polarizability tensor 
$\underline{\alpha}_{ij}(\tau)=(i/\hbar) \theta(\tau) \mathrm{tr} \{ [ \hat{\bf d}_i(\tau), \hat{\bf d}_j(0)] \hat{\rho}_{\rm KMS} \}$
%
\begin{equation}
\underline{S}(\omega)= \frac{\hbar}{\pi}  \theta(\omega) \underline{\alpha}_I(\omega),
\label{FDT}
\end{equation}
where $\theta(\omega)$ is the step function and $\underline{\alpha}(\omega)$ is the Fourier transform of $\underline{\alpha}(\tau)$.
Equation \eqref{FDT} is valid for the two previous models for the atom, since the
equilibrium FDT holds not only for linear but also for non-linear systems \cite{Polevoi75,Weiss08}, including an atom treated using a (nonlinear) two- or multi-level model. This can be seen in the following derivation of the FDT, showing its validity for an arbitrary (time independent) system Hamiltonian
$\hat{H}$ \cite{Kubo66,Talkner86}. Let $\hat{A}$ and $\hat{B}$ be two observables, and define
$M_{AB}(\tau) = \langle \hat{A}(\tau) \hat{B}(0) \rangle - \langle \hat{A}(0) \rangle \langle \hat{B}(0) \rangle$ and
$\chi_{AB}(\tau) = (i/\hbar) \langle [ \hat{A}(\tau), \hat{B}(0)] \rangle$. Then
$\chi_{AB}(\tau) =(i/\hbar) (M_{AB}(\tau) - M_{BA}(-\tau))$. For $\alpha_{AB}(\tau) = \theta(\tau) \chi_{AB}(\tau)$,
it follows that 
$\alpha_{AB}(\omega)-\alpha^*_{BA}(\omega) = (i/\hbar)\int_{-\infty}^{\infty} d\tau e^{i \omega \tau} [M_{AB}(\tau) - M_{BA}(-\tau)]$.
Using the equilibrium KMS condition $M_{BA}(-(\tau+i \hbar \beta)) = M_{AB}(\tau)$ \cite{Kubo57,Martin59}, 
we have
\begin{equation}
S_{AB}(\omega) = \frac{\hbar}{2 \pi i (1-e^{-\beta \hbar \omega})} [ \alpha_{AB}(\omega) - \alpha^*_{BA}(\omega) ] ,
\end{equation}
which  reduces to \eqref{FDT} in our case.  
Both for the oscillator and the two-level atom,  $\underline{C}$ and $\underline{\alpha}$ are symmetric tensors, and therefore the power spectrum $\underline{S}(\omega)$ is real.
Note that  $ \underline{\alpha}$ is the non-perturbative polarizability that depends on the optical properties of the
surrounding field, the atom, and the surface, and is a function of the atom's position ${\bf r}_a$ (omitted in the following for simplicity). 
Taking the large-time limit of (\ref{CPforce}) and using the FDT, one obtains the  (non-perturbative and non-Markovian) Casimir-Polder force \cite{Casimir48a}
\begin{equation}
F_{\rm CP} = \frac{\hbar}{\pi} \int_0^{\infty} d\xi \; {\rm Tr} 
\{  \underline{\alpha}(i \xi)   \cdot \partial_{z_a}  \underline{G}({\bf r}_a, {\bf r},i \xi) |_{{ \bf r} = {\bf r}_a} 
\} .
\label{CP-exact}
\end{equation}


Another commonly used fluctuation relation is the regression theorem \cite{Onsager31} and its generalization to the quantum case, known as the quantum regression hypothesis (sometimes called
``theorem") given by the Lax formula \cite{Lax63}. The quantum regression theorem (QRT) is  approximate, valid only in the weak system-bath coupling limit and near a resonance (see, for example, \cite{Talkner86,Ford96a}).
Although successfully used in quantum optics within its range of validity, the QRT is known to fail whenever non-Markovian and off-resonance effects play an important role \cite{Ford00}: the broadband nature of fluctuation-induced interactions suggests that its use in this context is therefore questionable.
Within the QRT the two-time dipole correlation tensor for a two-state atom or a harmonic oscillator for $t\to\infty$ is given by
$\underline{C}_{ij}(t,t-\tau)= {\bf d}_i {\bf d}_j e^{-i ( \omega_a - i  \gamma_a/2)\tau}$, where $\omega_a$ and $\gamma_a$ are the atomic transition frequency and dissipation rate, respectively.
Using this expression in (\ref{CPforce}) and taking the large time limit one obtains a Casimir-Polder force of the same form as (\ref{CP-exact}), but with $\underline{\alpha}(i \xi)$ replaced by $[ \underline{\tilde{\alpha}}(i \xi) +\underline{\tilde{\alpha}}(-i \xi)]/2$, where
$\underline{\tilde{\alpha}}_{ij}(i \xi) = ({\bf d}_i {\bf d}_j / \hbar) [ (\omega_a - i \xi - i \gamma_a/2)^{-1} +(\omega_a +i \xi + i \gamma_a/2)^{-1}]$ 
is the generalized ground state atomic polarizabilty \cite{Buhmann04}.
The QRT fails to give the expression (\ref{CP-exact}) predicted by the FDT and the exact solution for the harmonic oscillator model \cite{Intravaia11}, which coincides with \eqref{CP-exact} and reduces to the well-known Lifshitz formula.

The mathematical reason for this discrepancy lies in the distinct large-time behavior of the correlation tensor $\underline{C}(\tau)$. While the QRT predicts an exponential decay, the exact FDT results in a power-law decay for large times $\tau \gamma_a \gg 1$ (and agrees with the QRT only for
$\gamma_a \tau \lesssim 1$). For example, in the large time limit,  $\underline{C}(\tau) \propto \tau^{-2}$ for $\alpha_I(\omega) \propto \omega$ (Ohmic dissipation). 
Only in the weak coupling limit ($\gamma_a \rightarrow 0$), corresponding to a second-order perturbative calculation in powers of the coupling strengths ${\bf d}$, does the QRT coincide with the FDT.
A related phenomenon takes place in the spontaneous decay of an excited atom in vacuum, which in the Wigner-Weisskopf approximation is predicted to be exponential, but has large-time power-law corrections \cite{Berman10}.

\begin{figure}[t]
\includegraphics[width=7.5cm]{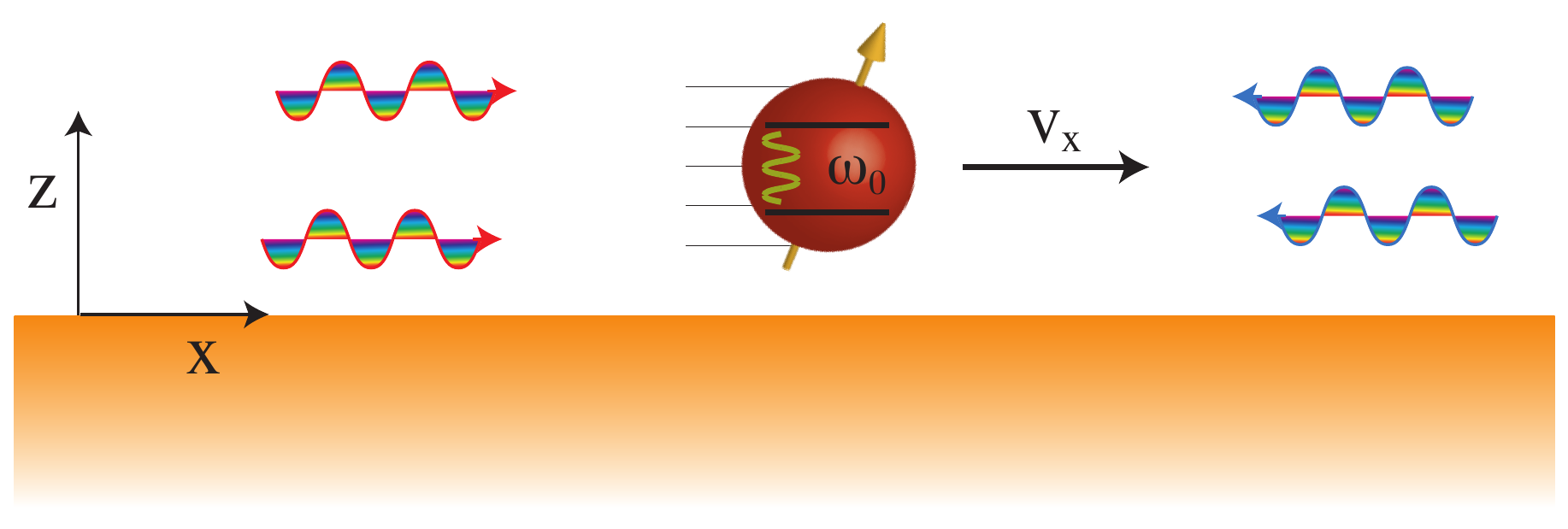}
\vspace{-.3cm}
\caption{Quantum friction on an atom moving at constant velocity above a surface}
\label{friction}
\end{figure}

The previous analysis shows that, beyond the weak coupling regime, the correct large time behavior of the 
two-time correlation tensor strongly affects the steady state Casimir-Polder force in (\ref{CPforce}). 
We show now that similar considerations also apply to the non-equilibrium situation of  an atom moving parallel (along the $x$-direction) to a flat semi-infinite ($z\le0$) bulk (Fig.\ref{friction}).  As before, we model the atom by an electric dipole operator and treat its center-of-mass coordinate ${\bf r}_a(t)$ semiclassically. The quantum frictional force is given by 
$F_{\rm fric}(t) = \langle \hat{\bf d}(t)\cdot\partial_{x_a} \hat{\bf E}({\bf r}_a(t),t) \rangle$,
where the expectation value is taken with respect to an initial uncorrelated 
atom+field/matter state in which the field/matter is in its vacuum state \cite{footnote}.
The $x$-dynamics is governed by $m_a \ddot{x}_a(t)=F_{\rm ext}(t)+ F_{\rm fric}(t)$, where $F_{\rm ext}(t)$ is an external classical force on the atom that drives it from the initial rest state at ${\bf r}_a(t=0)=(x_a,y_a,z_a)$ 
to a steady-state at time $t_s$ after which the atom moves at constant
velocity $v_x$ above the surface, ${\bf r}_a(t) = (x_a+v_x t, y_a,z_a)$. 
In the large-time limit, the stationary frictional force is given by
\begin{multline} 
F_{\rm fric} =- 
\mathrm{Re} \left\{ \frac{2}{\pi}\int\frac{d^{2}\mathbf{k}}{(2\pi)^{2}}  k_{x} \int_{0}^{\infty}d\omega  \int_{0}^{\infty}d\tau  \right.
\\
\left. \times e^{-i  (\omega-k_{x}v_{x})\tau} \;  \mathrm{Tr}\left[\underline{C}(\tau;v_{x})\cdot 
\underline{G}_{I}(\mathbf{k},z_a,z_a,\omega)\right] \right\}.
\label{fxsteady}
\end{multline}
Here $\underline{C}_{ij}(\tau; v_x) = {\rm tr} \{ \hat{\bf d}_i(\tau) \hat{\bf d}_j(0) \hat{\rho}(\infty) \}$ is the two-time correlation tensor in the 
non-equilibrium stationary state $\hat{\rho}(\infty)$ of the coupled moving atom plus field/matter. Note that it depends on the velocity of the atom, which is denoted by the $v_x$ dependency after the semi-colon in the expression above. Once more, we emphasize that
$\hat{\bf d}(\tau)$ contains the exact dynamics of the moving atomic dipole, i.e. including the backaction from the field/matter.

There is an extensive literature on non-equilibrium fluctuation theorems,
trying to generalize fundamental equilibrium results such as the fluctuation-dissipation theorem to non-equilibrium steady-state configurations (see, for example, \cite{Chetrite08,Baiesi09}).
One of the challenges is to identify the form of the non-equilibrium stationary density matrix, which is no
longer described by a KMS state but is model-dependent.  Despite this limitation, 
we will show that it is still possible to draw general conclusions about the frictional force in the low velocity limit.
In analogy to the static case, we define a power spectrum 
$\underline{S}(\omega; v_x) = (2 \pi)^{-1} \int_{-\infty}^{\infty} d\tau e^{i \omega \tau} \underline{C}(\tau; v_x)$, 
which is again a real and symmetric tensor since in our description $\underline{C}$ is symmetric. 
Using the symmetry properties of the Green tensor $\underline{G}$ for the homogeneous planar surface (see \cite{Wylie84}, for example),  (\ref{fxsteady}) can be re-written as
\begin{multline}
F_{\rm fric} =-
2 \int  \frac{d^2 {\bf k}}{(2\pi)^{2}} k_x \int_{0}^{\infty} d\omega  
\label{friction2} 
\\ \times \mathrm{Tr}\left[\underline{S}(k_{x}v_{x}-\omega; v_{x})\cdot \underline{G}_{I}(\mathbf{k},z_a,z_a,\omega)\right] .
\end{multline}
Note that in this expression the power spectrum $\underline{S}$ depends on the wave vector only through the Doppler shifted frequency  
$\omega-v_x k_x$. The friction is the momentum transfer $\hbar k_x$ to the atom weighted by
its Doppler-shifted power spectrum and by the electromagnetic density of states, all integrated over frequency and momentum.
As expected, the force vanishes for $v_x=0$: since $\underline{S}$ is symmetric
only the symmetric part of $\underline{G}_I$ (even in $k_{x}$  \cite{Wylie84}) is relevant. The integral then vanishes for parity reasons. 

Generally one is interested in computing $F_{\rm fric}$ to leading-order in $v_{x}$. For this, however,  one needs to know the expression for 
$S(\omega;v_x)$, which in general is not available (see, however, the harmonic oscillator model below). Nevertheless,
even without this knowledge, it is possible to prove that at zero temperature and in the stationary limit ($t \rightarrow \infty$)
there are no linear in $v_x$ terms in the friction force, independently
of the model for the atom's polarizability. Indeed, terms proportional to $v_x$ could only arise either from 
$\underline{S}(-\omega;v_x)$ or from $\underline{S}(k_x v_x - \omega; 0)$. 
The contribution of the former term cancels again for parity reasons upon integration over $k_x$.
The latter term, corresponding to a stationary state $\hat{\rho}(\infty)$ in which the atom is static, can be evaluated using the equilibrium FDT (\ref{FDT}), i.e. $\underline{S}(k_x v_x - \omega;0)= (\hbar/\pi)  \theta(k_x v_x - \omega) \underline{\alpha}_I(k_x v_x - \omega)$. 
Because of the motion-induced Doppler-shift, only frequency modes $0 \le \omega \le k_{x}v_{x}$ contribute, implying that very low frequencies are
relevant at small velocities. Since the atomic polarizability and the Green tensor are susceptibilities, they satisfy the crossing relation and their imaginary parts, being odd in $\omega$, vanish at $\omega=0$ in our case \cite{Landau80a}.  An expansion for small $v_{x}$ leads then to
\begin{eqnarray}
F_{\rm fric} & \approx & -  \frac{2 \hbar v_x^3}{3 (2 \pi)^3} \int_{-\infty}^{\infty} dk_y \int_0^{\infty} dk_x k_x^4 {\rm Tr}
[\underline{\alpha}_I'(0) \cdot \underline{G}'_I({\bf k},0) ]  \nonumber \\
& \approx & - \frac{45 \hbar v_x^3}{256 \pi^2 \epsilon_{0} z_a^7} \alpha'_I(z_a,0) \Delta'_I(0) ,
\label{friction-exact}
\end{eqnarray}
where in the first line we omitted to write the $z_a$ dependency of the Green tensor at coincidence.
In the second line  we have considered the low-frequency (near-field) form of the Green tensor for a dielectric
semi-space described by a complex permittivity $\epsilon(\omega)$ ($\epsilon_{0}$ in the vacuum permittivity), with 
$\Delta(\omega) \equiv  [\epsilon(\omega)-1]/[\epsilon(\omega)+1]$, and we have used 
$\underline{\alpha}(z_{a},\omega) = \delta_{ij} \alpha(z_{a},\omega)$ (we have reintroduced $z_a$ to underscore the dependency
of the dressed polarizability on the position of the atom).
The above argument proves that, within our description for the atom, the lowest-order expansion in velocity of the zero-temperature, stationary frictional force on an atom moving above a planar surface is at least cubic in $v_x$. In principle, however, in addition to that in \eqref{friction-exact} there could be other $v_x^3$ contributions to the frictional force arising from $v_x$-derivatives of $S(k_x v_x-\omega;v_x)$. Also, when either of the $\omega$-derivatives of the two tensors in (\ref{friction-exact}) vanish at $\omega=0$, higher-order terms in $v_x$ must be considered.

Regarding the dependency of the stationary frictional force (\ref{friction-exact}) on the atom-surface
separation, it must be emphasized that the $z_a^{-7}$ scaling arises solely from the $z_a$-dependency of the Green tensor.
In addition, as explained above, the power spectrum $\underline{S}$ and the polarizability $\underline{\alpha}$ implicitly
depend on $z_a$ via the exact dynamics of the coupled atom-field/matter system. In particular, these quantities are related to the atomic decay,
which at short distances and to lowest order in perturbation theory scales as $z_a^{-3}$, leading in (\ref{friction-exact}) to a total $z_a^{-10}$ dependency of the frictional force. For systems with intrinsic dissipation (e.g. gold nanoparticles) the radiation-induced damping is generally negligible and the frictional force has therefore a milder dependency on separation \cite{Volokitin07}.

In contrast to the FDT, the QRT predicts that for slow velocities the quantum frictional force is linear in $v_x$. 
As shown above, such a dependency results in principle from contributions of $S(k_x v_x -\omega;0)$ in 
(\ref{friction2}). Using the QRT expression for the two-time correlation tensor in the static case, 
$\underline{C}_{ij}(t,t-\tau; 0)= {\bf d}_i {\bf d}_j  e^{-i ( \omega_a - i  \gamma_a/2)\tau}$, and taking
the $t \rightarrow \infty$ limit one obtains indeed at the leading order expansion
\begin{eqnarray}
&& F_{\rm fric}^{\rm QRT} \approx  
v_x \frac{2 |{\bf d}|^2  \gamma_a}{3 \pi} 
\int \frac{d^2{\bf k}}{(2 \pi)^2} k_x^2 \int_0^{\infty} d\omega \nonumber \\
&& \times \frac{\omega+\omega_a} {[(\omega+\omega_a)^2+\gamma_a^2/4]^2}
{\rm Tr} [\underline{G}_{I}(\mathbf{k},z_a,z_a, \omega)] ,
\label{QRT-friction}
\end{eqnarray}
where, for simplicity, we assumed that the atom is isotropic  \cite{Scheel09}.
As for the static Casimir-Polder force, the quantum regression hypothesis fails to give the correct quantum frictional
force. Note, however, that once again both the FDT and the QRT give the same quantum frictional force in the limit $\gamma_a \rightarrow 0$, consistent with the observation before that the quantum regression hypothesis coincides with the exact fluctuation-dissipation theorem for systems near equilibrium in the weak coupling limit.  
In this limit, the resulting force is exponentially suppressed in $v_x^{-1}$ \cite{Dedkov02,Barton10b}. 
Linear response relations in fluctuational electrodynamics, based on equilibrium fluctuations, can also be employed to study quantum friction for small perturbations  around the equilibrium state 
\cite{Volokitin02,Dedkov02,Zurita-Sanchez04,Pieplow13,Maghrebi13,Golyk13}. 
In agreement with our analysis, at zero temperature the linear-in-velocity frictional force vanishes. However, far from equilibrium situations require fully non-equilibrium fluctuation relations.

The previous derivation uses general principles based on the fluctuation-dissipation theorem in non-equilibrium settings. 
In the following, we present an alternative derivation that does not resort to the FDT, and compute quantum friction for the harmonic oscillator model by directly solving
the equations of motion for the atomic dipole in the stationary limit (in the Supplemental Material we present a similar derivation for the two-state system). 
The dynamics of the dipole operator for the moving harmonic oscillator atom can be solved for exactly. Its equation of motion, including the back reaction of the electromagnetic field, is given by
$\ddot{\hat{q}}(t)+ \omega_a^2 \hat{q}(t) = (2\omega_{a}/\hbar){\bf d}\cdot \hat{\bf E}({\bf r}_a(t),t)$. 
Splitting the solution to Maxwell's equations for the total field $\hat{\bf E}$ as a sum of free ($\hat{\bf E}_0$, homogeneous solution) and source ($\hat{\bf E}_S$, particular solution) parts and
taking the Fourier transform, the equation of motion can be re-written as
$[-\omega^2 + \omega_a^2 -\frac{2\omega_{a}}{\hbar} \int \frac{d^2{\bf k}}{(2 \pi)^2} {\bf d} \cdot \underline{G}({\bf k},z_a,z_a,\omega+k_x v_x)
\cdot {\bf d} ] \hat{q}(\omega) = \frac{2\omega_{a}}{\hbar}
 \int \frac{d^2{\bf k}}{(2 \pi)^2} {\bf d} \cdot \hat{\bf E}_0({\bf k},z_a,\omega+k_x v_x) e^{i (k_x x_a + k_y y_a)}$. 
The polarizability of the moving oscillator is then given by
$\underline{\alpha}_{ij}(\omega;v_{x})=\frac{2\omega_{a}}{\hbar}{\bf d}_i {\bf d}_j
\left[ -\omega^2 + \omega_a^2 -\frac{2\omega_{a}}{\hbar} \int \frac{d^2{\bf k}}{(2 \pi)^{2}} {\bf d} \cdot  \underline{G}({\bf k},\omega+k_x v_x) \cdot 
{\bf d} \right]^{-1}$, where we have omitted to write the $z_a$ dependency of the Green tensor. The dynamic power spectrum $\underline{S}(\omega;v_x)$ is computed starting from $\langle \hat{\bf d}_i(\omega) \hat{\bf d}_j(\omega') \rangle= {\bf d}_i {\bf d}_j \langle \hat{q}(\omega) \hat{q}(\omega') \rangle$ and using that 
$\underline{S}_{ij}(\omega;v_x) = \frac{1}{2\pi} \int_{-\infty}^{\infty} \frac{d\omega'}{2 \pi} \langle \hat{\bf d}_i(\omega) \hat{\bf d}_j(\omega') \rangle$. 
The resulting exact expression for the zero-temperature case is
\begin{equation} 
\underline{S}(\omega;v_{x}) =\frac{\hbar}{\pi}  \theta(\omega) \underline{\alpha}_I(\omega;v_x)  -  \frac{\hbar}{\pi}  \underline{J}(\omega; v_{x}),
\label{non-eq-FDT}
\end{equation}
where the ``current"  $\underline{J}$  is given by
\begin{eqnarray}
&& \underline{J}(\omega;v_x) = \int \frac{d^2{\bf k}}{(2\pi)^2} 
[\theta(\omega) - \theta(\omega+k_x v_x)] \nonumber \\
&& 
\times \underline{\alpha}(\omega;v_x)  \cdot  \underline{G}_I({\bf k},\omega+k_x v_x) \cdot \underline{\alpha}^*(\omega;v_x) .
\end{eqnarray}
Generalized FDT relations for non-equilibrium, stationary classical systems \cite{Chetrite08} have the same structure as (\ref{non-eq-FDT}).  
Since only the symmetric part of the Green tensor contributes to ${\bf d} \cdot  \underline{G}({\bf k},\omega+k_x v_x) \cdot {\bf d}$, from the previous expressions for the polarizability and the current $\underline{J}$, we can deduce that the power spectrum is even in $v_{x}$. 
Using the identity $\underline{\alpha}_{I}(\omega;v_x) = \int \frac{d^{2}\mathbf{k}}{(2\pi)^{2}}\underline{\alpha}(\omega;v_x)  \cdot  \underline{G}_{I}(\mathbf{k},\omega+k_{x}v_{x})\cdot \underline{\alpha}^*(\omega;v_x)$, we rewrite the power spectrum \eqref{non-eq-FDT} as $\underline{S}(\omega;v_{x})=\frac{\hbar}{\pi} \int \frac{d^{2}\mathbf{k}}{(2\pi)^{2}}\theta(\omega+k_x v_x) \, \underline{\alpha}(\omega,v_{x})\cdot \underline{G}_{I}(\mathbf{k},\omega+k_{x}v_{x})\cdot \underline{\alpha}^*(\omega,v_{x})$.
An expansion at low velocity takes the form 
\begin{equation}
\underline{S}(\omega;v_{x})\approx \frac{\hbar}{\pi} \theta(\omega)\left[\underline{\alpha}_{I}(\omega;0)+\underline{\eta}(\omega;0) \frac{v_{x}^{2}}{2} \right]+\mathcal{O}(v_{x}^{4}) .
\label{Sexpansion}
\end{equation}
Here we have defined
$\underline{\eta}(\omega;0) =
\underline{\alpha}^{''}(\omega;0)\cdot\underline{G}_{I}(\omega)\cdot\underline{\alpha}^{*}(\omega;0) +
\underline{\alpha}(\omega;0)\cdot\underline{g}(\omega)\cdot\underline{\alpha}^{*}(\omega;0) +
\underline{\alpha}(\omega;0)\cdot\underline{G}_{I}(\omega)\cdot[\underline{\alpha}^{''}(\omega;0)]^{*}$ (the double prime denotes
second derivative with respect to velocity), and $\underline{g}(\omega) = \int d^2{\bf k} (2 \pi)^{-2} k_x^2 \partial^2_{\omega} \underline{G}_I({\bf k},\omega)$.
The tensor $\underline{\eta}(\omega;0)$ vanishes at $\omega= 0$ because  it is a sum of terms proportional either to the imaginary part of the Green tensor or to its second derivative.  Using \eqref{Sexpansion} in \eqref{friction2} one can verify that to leading order in $v_x$ the quantum frictional force for the harmonic oscillator model is 
exactly given by \eqref{friction-exact}, and the next order is proportional to $v_x^5$ (see Supplemental Material). 

Our result for the $v_x^3$ dependence of the quantum friction force on a moving atom contrasts with some previous
works in the literature that predicted a zero-temperature frictional force linear in $v_x$. In \cite{Scheel09} the atom was modeled as a multi-level
system and the dipole correlation function in (\ref{fxsteady}) was computed using QRT, which lead to a stationary friction force linear
in velocity (\ref{QRT-friction}).  Calculations of quantum friction based on QRT, Wigner-Weisskopf, or Markovian approximations encompass
an exponential-only decay of the dipole correlation tensor, which is valid for times $t \lesssim \gamma_a^{-1}$. Importantly, they miss the power-law decay at larger times  $t \gg \gamma_a^{-1}$ which strongly affects the low-frequency behavior of the spectrum. The discussion after (\ref{friction2}) shows that, in the stationary case, quantum friction is a low-frequency
phenomenon (see also the paragraph after \eqref{CP-exact}). Therefore, it is not surprising that the above mentioned approximations fail to predict the correct stationary behavior and lead to a different dependence
of the force on the atom's velocity. On the other hand, in \cite{Barton10b}  the atom was modeled as a harmonic oscillator and, by calculating the power dissipated by the atom into pairs of surface plasmons using an approach based on standard perturbation theory, a linear-in-velocity frictional force similar to \cite{Scheel09}  was obtained (within the same approximations an identical result is obtained for a two-level atom).
This time-dependent perturbative approach assumed that the atom
remains in its bare ground state, and is valid for times not too long for which decays are still exponential. In contrast, our previous
calculation shows that in the large-time, non-equilibrium steady-state the quantum frictional force becomes cubic in velocity.

Due to the weak nature of quantum friction, its experimental detection is challenging. Indeed, in the near field our result (\ref{friction-exact}) takes the form 
\begin{equation}
F_{\rm fric} \approx  - \frac{90}{\pi^3}\frac{\hbar \rho^{2}\alpha^{2}_{0}}{(2 z_{a})^{10}} v_x^3 ,
\end{equation}
where $\rho$ is surface's electrical resistivity and $\alpha_{0}$ the static atomic polarizability. As an example, for a ground state $^{87}$Rb atom 
($\alpha_0=5.26 \times 10^{-39} {\rm Hz}/ ({\rm V/m})^2$ \cite{Steck})
flying at $v_x=340$m/s at a distance  $z_a=10$ nm above a silicon semi-space ($\rho=6.4 \times 10^2 \Omega {\rm m}$), the zero temperature drag force is $F_{\rm fric} \approx - 1.3 \times 10^{-20}$N. Nevertheless, new experimental setups (e.g. new materials \cite{Volokitin11} and/or new geometries \cite{Intravaia13}) and techniques (e.g. atom-interferometry) could make it accessible in the near future.

In summary, we have studied quantum friction using general concepts of quantum statistical mechanics.
We have derived a generalized  non-equilibrium fluctuation-dissipation relation for an atom in steady motion above a surface, and shown that
at low speeds the quantum frictional force is cubic in velocity. The analysis can be extended to include thermal
fluctuations. In the high-temperature (classical) limit ($\hbar \beta \gamma\ll 1 $ \cite{Talkner86}), however, quantum regression agrees with the FDT \cite{Talkner86,Ford96a,Ford99}, and the resulting frictional force scales linearly with velocity. A study similar to the one present here can be 
performed for the case of macroscopic bodies in relative motion \cite{Volokitin07,Pendry97}). Finally, we would like to stress that our discussion of the implications and limitations of the use of fluctuation relations in calculations of equilibrium and non-equilibrium atom-surface interactions can potentially impact a broad range of fields such as atom interferometry and atom-chips. 

We are grateful to G. Barton, S. Buhmann, L. Cugliandolo,  J.P. Garrahan, C. Henkel, and S. Scheel for insightful discussions, and to the Alexander von Humboldt Foundation and the LANL LDRD program for financial support.


\begin{thebibliography}{10}
\bibitem{hamiltonian}
The two-state atom free Hamiltonian is $\hat{H}_{a}=(\hbar \omega_{a}/2)\hat{\sigma}_{3}$ where $[\hat{\sigma}_{i},\hat{\sigma}_{j}]=2\imath \varepsilon_{ijk}\hat{\sigma}_{k}$. In the harmonic oscillator case $\hat{H}_{a}=(\hbar\omega_{a}/4)\left(\hat{p}^{2}+\hat{q}^{2}\right)$ with $[\hat{q},\hat{p}]=2 i$.
\bibitem{footnote}
We are not including the R\"{o}entgen interaction $- \hat{\bf d} \cdot {\bf v} \times \hat{\bf B}$ which, for the purpose of calculating
the friction force, is negligible as compared to the nonretarded interaction $- \hat{\bf d} \cdot \hat{\bf E}$, as shown in \cite{Scheel09}.
\end{thebibliography}

\section*{\large Supplementary information}

Here, we compute quantum friction on a moving two-level atom by solving the equation of motion for the atomic dipole in perturbation theory.
The dynamics of a two-state system can be derived from the Hamiltonian
\begin{equation}
\hat{H}=\frac{\hbar \omega_{a}}{2}\hat{\sigma}_{3}+\hat{H}_{\rm field}- \mathbf{d}\cdot \hat{\bf E}(\mathbf{r}_{a})\hat{\sigma}_{1} ,
\label{hamiltonian}
\end{equation}
where $\hat{\sigma}_{3}=|e\rangle\langle e|-|g\rangle\langle g|$ and $\hat{\sigma}_{1}=|e\rangle\langle g|+|g\rangle\langle e|$; together with $\hat{\sigma}_{2}=i(|g\rangle\langle e|-|e\rangle\langle g|)$ they satisfy the algebra of Pauli matrices. $\hat{H}_{\rm field}$ is the free electromagnetic field Hamiltonian.
The internal state dynamics  is given by 
\begin{equation}
\ddot{\hat{\sigma}}_{1}(t)+\omega^{2}_a \hat{\sigma}_{1}(t)=- (2\omega_a/ \hbar)  \hat{\sigma}_{3}(t) {\bf d} \cdot \hat{\bf E}(\mathbf{r}_{a}(t),t).
\label{exactTSA}
\end{equation}
This is a nonlinear equation that does not allow for an exact solution, and in the following we solve it using a perturbative scheme in powers of the dipole coupling
${\bf d}$. 

The computation of the quantum frictional force requires the evaluation of the two-time correlator tensor  
$\underline{C}_{ij}(t,t';v_x) =  {\bf d}_i {\bf d}_j \langle \hat{\sigma}_1(t) \hat{\sigma}_1(t')\rangle$. To second-order it can be evaluated from the Pauli matrices' free evolution,
and at zero temperature and for a ground state atom $\underline{C}_{ij}={\bf d}_i {\bf d}_j e^{-i \omega_a (t-t')}$, resulting in a frictional force that is exponentially suppressed in $v_x^{-1}$. To compute the frictional force to fourth-order one needs to evaluate $\langle \hat{\sigma}_1(t) \hat{\sigma}_1(t')\rangle$ at second order. To this end we first insert in equation \eqref{exactTSA} the formal solution for the dynamics of  $\hat{\sigma}_{3}(t)=\hat{\sigma}_{3}(0)+\frac{2}{\hbar\omega_{a}}\int_{0}^{t}dt_{1} \dot{\hat{\sigma}}_{1}(t'){\bf d}\cdot\mathbfh{E}(\mathbf{r}_{a}(t_{1}),t_{1})$, and then replace the exact field ${\bf E}({\bf r},t)$ by its free evolution ${\bf E}_0({\bf r},t)$. At second order we obtain
\begin{multline}
\ddot{\hat{\sigma}}_{1}(t) + \frac{2}{\hbar^2} \int_{0}^{t}dt_{1} \{\mathbf{d}\cdot \hat{\mathbf{E}}_{0}(\mathbf{r}_{a}(t),t),\mathbf{d}\cdot \hat{\mathbf{E}}_{0}(\mathbf{r}_{a}(t_{1}),t_{1})\}\dot{\hat{\sigma}}_{1}(t_{1}) \\
+ \omega^{2}_{a} \hat{\sigma}_{1}(t) = - \frac{2\omega_a}{\hbar} \hat{\sigma}_{3}(0)\mathbf{d}\cdot \hat{\mathbf{E}}_{0}(\mathbf{r}_{a}(t),t).
\end{multline} 
Multiplying this equation from the right by $\hat{\sigma}_1(t')$, averaging on the initial factorized state, taking the infinite time limit, and finally
Fourier transforming the resulting equation, we can write the power spectrum to fourth order in the dipole coupling as (for simplicity, we omit the Green tensor's dependence on the position of the atom)
%
%
\begin{multline}
\underline{S}(\omega,v_{x}) = \frac{\hbar}{\pi}\int \frac{d^{2}\mathbf{k}}{(2\pi)^{2}}\theta(\omega+k_x v_x) \\
\times\underline{\alpha}(\omega;v_{x})\cdot \underline{G}_{I}(\mathbf{k},\omega+k_{x}v_{x})\cdot\underline{\alpha}^{*}(\omega;v_{x})
\label{TSA_S}
\end{multline}
where  $\underline{\alpha}(\omega;v_{x})=(2\omega_{a}/\hbar){\bf d}{\bf d}[\omega^{2}_{a}(1-\Delta)-\omega^{2}-\imath \omega\gamma]^{-1}$.  
The functions $\Delta(\omega;v_{x})=2\mathrm{P}\int_{0}^{\infty}\frac{d\omega'}{\pi} \frac{\omega^{2}}{\omega_{a}^{2}} \frac{\gamma(\omega',v_{x})}{\omega^{2}-\omega'^{2}}$ and $\gamma(\omega;v_{x})=\frac{2}{\hbar}\int\frac{d^{2}\mathbf{k}}{(2\pi)^{2}}\mathrm{sign}(\omega+k_{x}v_{x})\,\mathbf{d}\cdot \underline{G}_{I}(\mathbf{k},z_{a};\omega+k_{x}v_{x})\cdot\mathbf{d}$ are even in $\omega$ and
give the second order atomic (velocity dependent) frequency shift and decay rate. One can see that the dynamic power spectrum is symmetric and real.

We now study the low velocity expansion of the quantum frictional force on the two-level atom. We start by noting that
only the symmetric part of the Green tensor contributes to (\ref{TSA_S}), and since 
$\underline{G}^{\rm sym}(\mathbf{k},\omega)=\underline{G}^{\rm sym}(-\mathbf{k},\omega)$, the power spectrum
(\ref{TSA_S}) is even in $v_{x}$. Therefore, the power spectrum can be expanded as
$\underline{S}(\omega;v_{x})= \underline{S}(\omega;0)+\underline{S}^{''}(\omega;0)v_{x}^{2}/2+\mathcal{O}(v_{x}^{4})$,
where the double prime denotes second derivative with respect to velocity.
Defining 
$\underline{\tilde \alpha}_{I}(\omega;0) =\int \frac{d^2{\bf k}}{(2\pi)^{2}}  \underline{\alpha}(\omega;0)  \cdot  \underline{G}_I({\bf k},\omega) \cdot \underline{\alpha}^*(\omega;0)$,
we obtain low velocity expansion of the power spectrum of the two-level atom (valid to fourth order in the dipole coupling)
\begin{equation}
\underline{S}(\omega;v_x) \approx \frac{\hbar}{\pi} \theta(\omega) \left[ \underline{\tilde\alpha}_{I}(\omega;0) + \underline{\tilde\eta}(\omega;0) \frac{v_x^2}{2} \right]
+ {\cal O}(v_x^4) .
\label{TSAspectrumlowv}
\end{equation}
Note that it has the same form as the low velocity expansion of the harmonic oscillator model (see Eq. (11) of the main text). In this case, however, the function $ \underline{\tilde \alpha}_{I}(\omega)$ is  connected with the fourth order perturbative expression of the imaginary part of the polarizability while
$\underline{\tilde \eta}(\omega;0) =
\underline{\alpha}^{''}(\omega;0)\cdot\underline{G}_{I}(\omega)\cdot\underline{\alpha}^{*}(\omega;0) +
\underline{\alpha}(\omega;0)\cdot\underline{g}(\omega)\cdot\underline{\alpha}^{*}(\omega;0)
+\underline{\alpha}(\omega;0)\cdot\underline{G}_{I}(\omega)\cdot[\underline{\alpha}^{''}(\omega;0)]^{*}
$ (the double prime denotes
second derivative with respect to velocity), 
and $\underline{g}(\omega) = \int \frac{d^2{\bf k}}{(2 \pi)^{2}} k_x^2 \partial^2_{\omega} \underline{G}_I({\bf k},\omega)$.
We now use (\ref{TSAspectrumlowv}) in the expression for the quantum friction force (see Eq.(6) of the main paper)
\begin{multline}
F_{\rm fric} =-
2 \int  \frac{d^2 {\bf k}}{(2\pi)^{2}} k_x \int_{0}^{\infty} d\omega  
\label{friction2} 
\\ \times \mathrm{Tr}\left[\underline{S}(k_{x}v_{x}-\omega; v_{x})\cdot \underline{G}_{I}(\mathbf{k},z_a,z_a,\omega)\right] .
\end{multline}
We need to expand the integrals ($k_{x}>0$) 
\begin{multline}
I_1 = \int_0^{k_x v_x} d\omega {\rm Tr}  \left[ \underline{\tilde\alpha}_{I}(k_x v_x - \omega;0) \cdot \underline{G}_I(\mathbf{k},\omega) \right] ,  \\
I_2 = \frac{v_x^2}{2} \int_0^{k_x v_x} d\omega {\rm Tr}  \left[ \underline{\tilde\eta}(k_x v_x - \omega;0) \cdot \underline{G}_I(\mathbf{k},\omega) \right],
\end{multline}
to lowest order in $v_x$. Using that $\underline{\tilde \alpha}_{I}(\omega)$ and $\underline{G}_I(\mathbf{k},\omega)$
are odd in $\omega$, it follows that $I_1 \propto v_x^3$ and results in exactly the same quantum friction force as in Eq.(7) of the main text.
Using that also $\underline{\tilde\eta}(\omega)$ is odd in $\omega$, it follows that $I_2 \propto v_x^5$. 
Hence, the stationary quantum frictional on a moving two-level atom scales as the cubic power of its velocity to leading order, as
shown in the main paper using an alternative method based on the fluctuation-dissipation theorem.

\vspace{0.5cm}

\noindent F. Intravaia$^{1,2}$, R. O. Behunin$^{1,3,4}$, and D. A. R. Dalvit$^{1}$.
\begin{small}
\begin{enumerate}
\item[$^{1}$]
Theoretical Division, MS B213, Los Alamos National Laboratory, Los Alamos, New Mexico 87545, USA
\item[$^{2}$]
Max-Born-Institut, 12489 Berlin, Germany
\item[$^{3}$]
Center for Nonlinear Studies, Los Alamos National Laboratory, Los Alamos, New Mexico 87545, USA
\item[$^{4}$]
Department of Applied Physics, Yale University, New Haven, Connecticut 06511, USA
\end{enumerate}
\end{small}

\end{document}